\documentclass[conference]{IEEEtran}
\IEEEoverridecommandlockouts
\usepackage{amsmath,amssymb,amsfonts}
\usepackage{algorithmic}
\usepackage{graphicx}
\usepackage{textcomp}
\usepackage{xcolor}
\usepackage{bm}
\usepackage{subfigure}
\usepackage{multirow}
\usepackage{tabu}
\usepackage{makecell}
\usepackage[square, comma, sort&compress, numbers]{natbib}
\def\BibTeX{{\rm B\kern-.05em{\sc i\kern-.025em b}\kern-.08em
    T\kern-.1667em\lower.7ex\hbox{E}\kern-.125emX}}
\begin{document}

\title{Dual Learning-based Video Coding with Inception Dense Blocks }

\author{\IEEEauthorblockN{Chao Liu, Heming Sun\IEEEauthorrefmark{2}, Jun'an Chen, Zhengxue Cheng, Masaru Takeuchi, Jiro Katto, Xiaoyang Zeng and Yibo Fan\IEEEauthorrefmark{1}}
\IEEEauthorblockA{\IEEEauthorrefmark{1}State Key Lab of ASIC and System,
Fudan University, Shanghai P.R. China\\
Email: fanyibo@fudan.edu.cn}
\IEEEauthorblockA{\IEEEauthorrefmark{2}Waseda Research Institute for Science and Engineering, Waseda University, Tokyo Japan \\
Email: hemingsun@aoni.waseda.jp}
}
\maketitle

\begin{abstract}
In this paper, a dual learning-based method in intra coding is introduced for PCS Grand Challenge. 
This method is mainly composed of two parts: intra prediction and reconstruction filtering. 
They use different network structures, the neural network-based intra prediction uses the full-connected network to predict the block while the neural network-based reconstruction filtering utilizes the convolutional networks. 
Different with the previous filtering works, we use a network with more powerful feature extraction capabilities in our reconstruction filtering network. 
And the filtering unit is the block-level so as to achieve a more accurate filtering compensation. 
To our best knowledge, among all the learning-based methods, this is the first attempt to combine two different networks in one application, and we achieve the state-of-the-art performance for AI configuration on the HEVC Test sequences. 
The experimental result shows that our method leads to significant BD-rate saving for provided 8 sequences compared to HM-16.20 baseline (average 10.24\% and 3.57\% bitrate reductions for all-intra and random-access coding, respectively). 
For HEVC test sequences, our model also achieved a 9.70\% BD-rate saving compared to HM-16.20 baseline for all-intra configuration. 
\end{abstract}


\section{Introduction}
Recently, neural networks have shown great potential in various fields and many promising results have been achieved in video coding. Especially for intra coding \cite{HEVCIntra}, many creative ideas have been proposed to enhance the performance of intra coding. Intra coding uses the reference pixels to predict the current block, and the residual samples calculated by original pixels minus the predicted pixels are sent to transform and quantization to obtain the compressed residual. After the inverse quantization and inverse transform process, the residual samples with distortion is obtained. The reconstructed pixels at the decoding end can thus be obtained by adding the predicted pixels and the distorted residual samples.

Many neural network-based works are carried out for the prediction \cite{FC1, FC2, FC_CONV, HIT_CNN, RNN} and filtering \cite{ARCNN, VRCNN, RHCNN, MMSN}, which are two key parts of intra coding. In neural network-based intra prediction, network structures mainly consist of convolutional layers and full connected layers \cite{FC1, FC2, FC_CONV, HIT_CNN}. Li et al. \cite{FC1, FC2} use full connected network to explore the capacity of prediction. PNNS is introduced by Dumas et al. \cite{FC_CONV} based on both fully-connected and convolutional neural networks. Cui et al. \cite{HIT_CNN} propose IPCNN, and this network directly applies CNNs to intra prediction, which achieves a good performance in intra coding as well. And Hu et al. \cite{RNN} try Progressive Spatial Recurrent Neural Network (PS-RNN) and SATD loss function, which supports variable-block-size for intra prediction.

From a perspective of reconstruction filtering, similar convolutional neural network structures can be used for tasks such as super-resolution, denoising and filtering. Dong et al. \cite{SRCNN} proposed SR-CNN for super-resolution, and they also design AR-CNN \cite{ARCNN} for compression artifacts reduction based on it. Then Dai et al. \cite{VRCNN} increased the width of AR-CNN and proposed VR-CNN to further improve the network performance. In more complex network design, Zhang et al. \cite{RHCNN} proposed an RHCNN with 3,340,000 parameters, and achieved better experimental results. And a multi-modal/multi-scale model called MMS-net \cite{MMSN} with 2,298,160 parameters is proposed by Kang et al., which shows a multi-scale CNN structure can effectively improve image reconstruction performance.

For filtering tasks, we believe that VR-CNN does not adequately extract the full characteristics of the data. Deeper features can help guide the network for better filtering. While RHCNN is a little complicated for practical applications, and relatively simpler design needs to be proposed. Therefore, we have designed a new reconstruction filtering network with 475,233 parameters, which based on the inception networks\cite{inception4}. The inception network has an excellent performance in the classification task because of its powerful feature extraction capabilities. We believe that it can also perform well in filtering tasks. Specifically, the main contributions are as follows.
\begin{figure*}[tbp]
\centering
\subfigure[The network structure of VRCNN.]{
\begin{minipage}[t]{0.62\linewidth}
\centering
\includegraphics[scale=0.58]{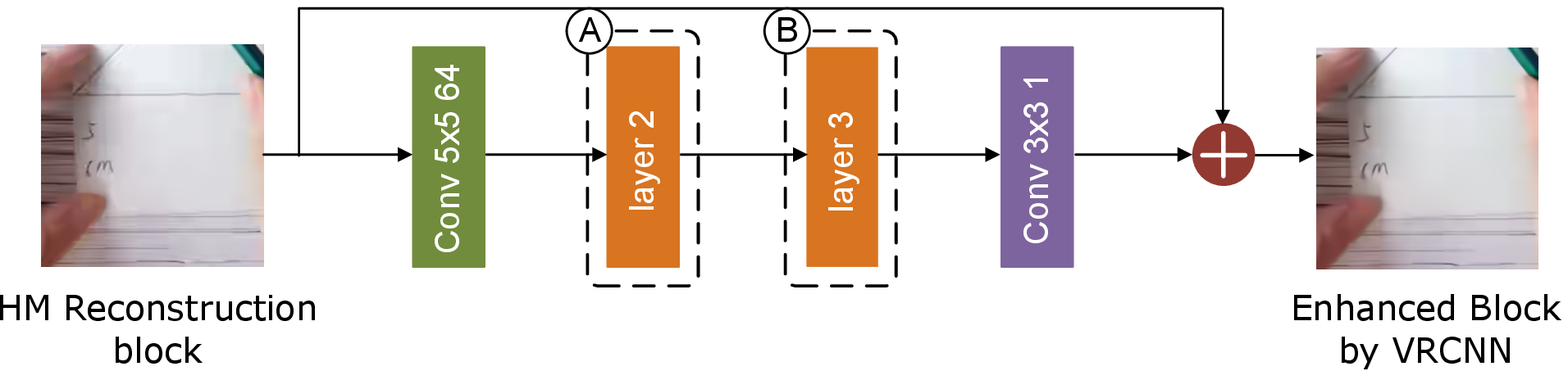}
\end{minipage}%
}%
\subfigure[The layer2 and layer3 of VRCNN.]{
\begin{minipage}[t]{0.38\linewidth}
\centering
\includegraphics[scale=0.58]{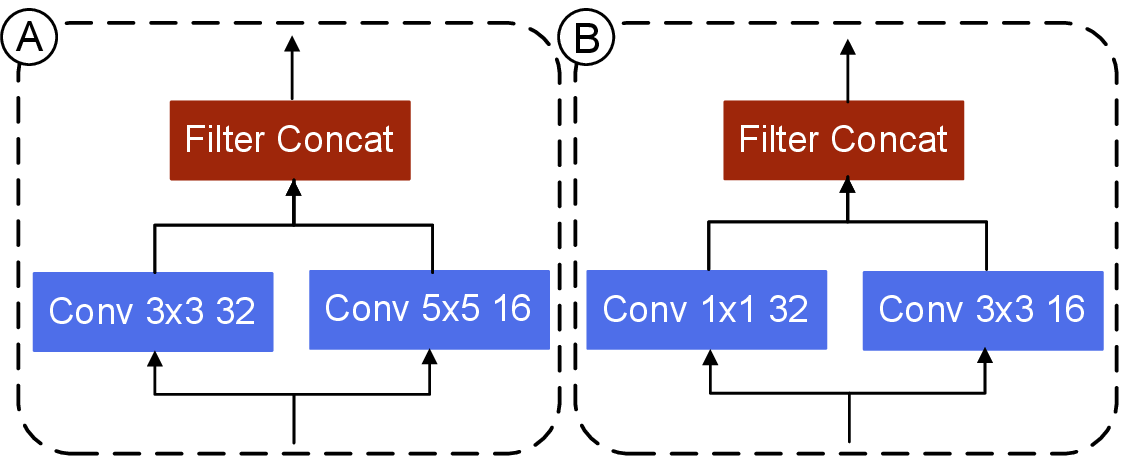}
\end{minipage}%
}%

\subfigure[The network structure of our reconstruction filtering.]{
\begin{minipage}[t]{0.62\linewidth}
\centering
\includegraphics[scale=0.58]{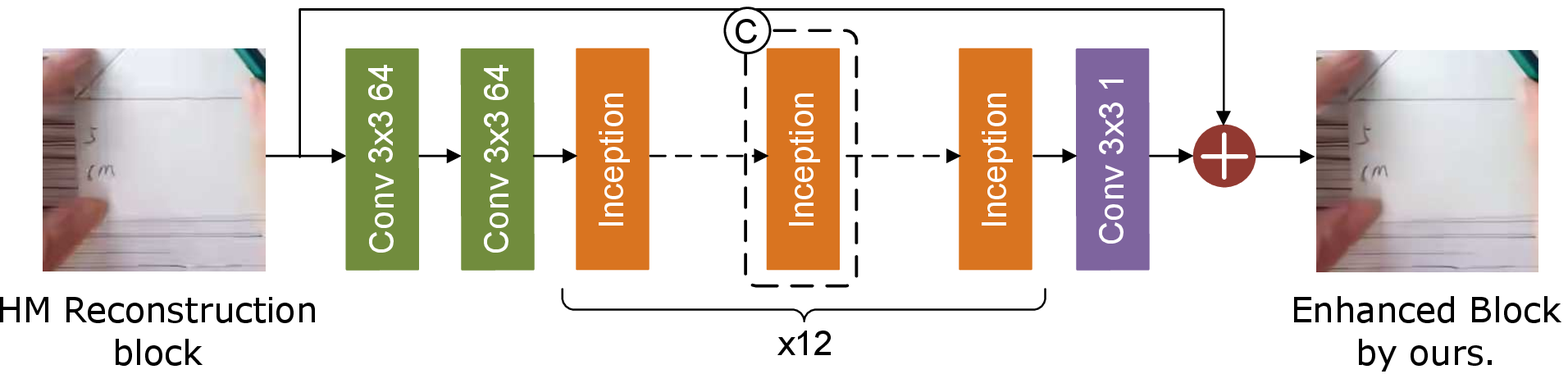}
\end{minipage}%
}%
\subfigure[The improved inception block.]{
\begin{minipage}[t]{0.38\linewidth}
\centering
\includegraphics[scale=0.58]{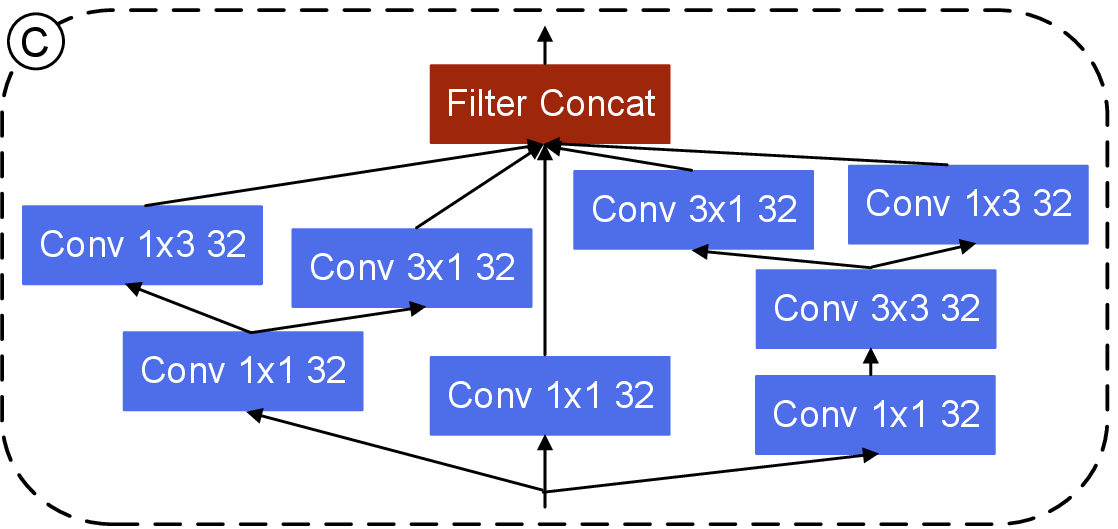}
\end{minipage}%
}%
\centering
\caption{The architecture of VR-CNN \cite{VRCNN} and our reconstruction filtering model.}\label{structure}
\end{figure*}

\begin{itemize}

\item In VR-CNN, the features extracted by two convolutional blocks of the two branches. We use the improved inception dense network to replace it so that the network has stronger feature extraction ability, which leads to a better filtering performance. After using inception filtering, we can achieve 3.07\%, 3.06\% and 3.17\% BD-rate improvement for YUV components, respectively. Moreover, the filtering network we are performing is not a frame-level but a block-level. Because block-level filtering enhance the quality of the reconstructed pixels, which directly increases the predictive performance and reduce the bit-rates of the residual in the next block.
\item We combine two learning-based methods and ensure that integrating two methods together can further improve the performance. The experimental results show that this integrating method achieves extra 2.42\% BD-rate saving for luminance component than using our reconstruction filtering model alone. It is thus found that the network has a superposition effect.
\end{itemize}

\section{Proposed Methods}
\subsection{Full-connected Network-based Intra Prediction}
For the intra prediction, originally, there are 35 modes in HEVC composed of two non-directional and 33 directional modes. Based on HEVC original modes, we append one neural network mode by using the fully-connected network in \cite{FC1, FC2}. The difference is that we apply the neural network mode at the prediction unit (PU) level. The mode signaling scheme is that one additional bin is consumed to represent whether the best mode of each PU is the neural network mode or not.
\subsection{Convolutional Network-based Reconstruction Filtering}
Like the structure of VRCNN in Fig.\ref{structure}(a), the reconstruction filtering network shown in Fig.\ref{structure}(c) consists of a pre-processed convolutional network shown in green rectangle, an intermediate inception dense network shown in orange rectangles and a post-processing convolutional network shown in the purple rectangle.

The first 64 feature maps convolutional layer with kernel size is 5x5 in VR-CNN pre-processed convolutional module (Fig.\ref{structure}(b)) is transformed into two convolutional layers of 64 feature maps with kernel size is 3x3, which helps to improve the extraction of basic features.
Compared to 2 blocks in VR-CNN, our middle part uses 12 improved inception network blocks to further extract the features. The specific structure is shown in Fig. Fig.\ref{structure}(d). Its input is the output from the front layer, and the chunk is composed of three branches, each branch has a 32 feature maps convolutional layers with kernel size is 1x1 as the first layer. In order to extract the features from different receptive fields, the two convolution layers with kernel size are 1x3 and 3x1 are connected to the first layer in one of the branches. In addition, another branch is connected to a convolution layer with a kernel size of 3x3 serially, and then connected with two convolution layers of 1x3 and 3x1 convolution kernels in parallel. Different with the original inception, we removed the pooling layer for a more compact network structure.
\begin{table*}[!tbp]
\small
  \centering
  \caption{The BD-rate results of proposed methods than HM in All-Intra and Random-Access for PCS grand challenge short videos }\label{com_pcs}
  \begin{tabu}{c|r|r|r|r|r|r|r|r|r|r}
    \Xhline{1.0pt}
    \multirow{3}*{Sequence}  & \multicolumn{5}{c|}{AI}& \multicolumn{5}{c}{RA} \rule{0pt}{9pt} \\
    \cline{2-11}
    ~& \multicolumn{3}{c|}{BD-rate}&\multirow{2}*{$\Delta T_{enc}$}&\multirow{2}*{$\Delta T_{dec}$}& \multicolumn{3}{c|}{BD-rate}&\multirow{2}*{$\Delta T_{enc}$}&\multirow{2}*{$\Delta T_{dec}$}  \\
    \cline{2-4}\cline{7-9}
    ~&\multicolumn{1}{c|}{Y}& \multicolumn{1}{c|}{U}&\multicolumn{1}{c|}{V}&~&~&\multicolumn{1}{c|}{Y}& \multicolumn{1}{c|}{U}&\multicolumn{1}{c|}{V}&~&~\\
    \hline
    01 &   -8.31\% & -14.51\% &  -13.22\%&7219\%&177661\%& -3.11\% & -8.56\% & -7.04\%&659\%&17880\% \\
    \hline
    02 &   -10.98\% & -14.27\% & -14.93\%&8449\%&233472\%&   -4.24\% & -3.82\% & -4.36\%&708\%&20839\% \\
    \hline
    03 &    -10.51\% & -5.46\% &  -14.80\%&8860\%&220486\%&   -3.65\% & -6.97\% & -3.47\%&478\%&11143\%\\
    \hline
    06 &   -11.40\% & -6.54\% & -10.00\%&8888\%&226057\%&   -2.99\% & -4.48\% & -1.57\%&584\%&11420\% \\
    \hline
    08 &  -7.91\% & -12.59\% & -12.35\%&7480\%&168358\%&   -2.11\% & -3.37\% & -3.09\%&636\%&10412\% \\
    \hline
    09 &   -8.63\% & -16.48\% & -17.29\%&6560\%&160278\%&   -2.12\% & -5.19\% & -5.46\%&623\%&11474\% \\
    \hline
    10 &    -11.40\% & -13.24\% & -13.55\%&9117\%&231279\%&   -3.24\% & -1.31\% & -2.20\%&594\%&10952\% \\
    \hline
    13 &  -12.83\% & -16.26\% & -17.77\%&7925\%&185475\%&   -7.09\% & -9.38\% & -9.74\%&540\%&10421\% \\
    \Xhline{0.8pt}
    Average & -10.24\% & -12.41\% & -14.24\%&8015\%&198322\%&   -3.57\% & -5.38\% & -4.61\%&599\%&12628\% \\
    \Xhline{1.0pt}
  \end{tabu}
\end{table*}
The last part is a post-processing convolution module that make the number of output feature maps return to the same number of the input. Because the inputs to our model have only one feature map, the last part is a convolutional layer with a kernel size of 3x3 and the number of feature maps is one.

The output of the post-processing convolution module and the reconstruction block will be added to obtain the filtered reconstruction block. Each convolutional layer except the last one is followed by an activation function "relu". And all convolutional layer padding methods use "same", which makes the input and output sizes unchanged.
The input to the designed network is 32x32 reconstructed block from HM. For a CTU with the size $64\times 64$, it can be divided into four 32x32 luminance component blocks and two 32x32 chroma component blocks with pixel format "YUV420". The output of the network is the enhanced reconstructed block. Because PSNR can reflect the performance of video coding relatively simply and effectively, we use the MSE (Mean Square Error) between enhanced reconstruction block $Y$ and the original reconstruction block $X$ as the loss function $L$ to train our network. The symbol $\Theta$ and $F$ are the parameters and outputs of our model respectively, and N is 32.
\begin{equation}\label{mse}
  L(\Theta)=\frac{1}{N^2}\sum_{i=1}^{N}\sum_{j=1}^{N}(F(Y(i,j)|\Theta)-X(i,j))^2
\end{equation}

\subsection{Block-level Filtering}
Different with the previous designs, our filter network is based on block-level rather than frame-level. After each CTU encoding is complete, the reconstruction blocks can be got on both the encoding end and the decoding end. They will be sent to the network to obtain the filtered result, thereby improve the image quality of the current block on the one hand, and provide more accurate reference pixels for the next block on the other hand. So this method can reduce the bitrates and improve the picture quality at the same time. Because our filter network is block-level, de-blocking may help to improve the edge of the block as well. Therefore it can coexist with the loop filter of the HEVC.

\section{Experimental Results}
\subsection{Experimental Setting}
The datasets we used is DIV2K \cite{DIV2K}, which consisted of 900 images with a resolution of 2k.
Because the reconstruction quality of the chrominance components is often too high, only the luminance component is used for training.
But in actual testing, chrominance components will also be tested by using our model. In order to avoid those blocks with accurate reconstruction are affected negatively by using neural network based filtering model. Different QP models are tested and finally the appropriate QPs for each target QP are selected while exporting the dataset. Specifically, we derive four datasets from different QPs to train four corresponding filtering models with different QP bands. 
%
The deep learning framework used is Keras \cite{keras} because of its better support for TensorFlow \cite{tensorflow}. We call the freeze model of TensorFlow converted from Keras model in the actual application.
The GPUs used for training and the CPUs used for testing were NVIDIA GeForce RTX 2080 and Intel Xeon Gold 6134 at 3.20 GHz, respectively.

\subsection{Comparison with HEVC Baseline}

We use the 8 PCS grand challenge short videos to test, and the test condition includes four QPs (22 27 32 37) and two configurations (AI and RA). The test results are shown in TABLE \ref{com_pcs}. In the AI configuration, for the luminance component, we get BD-rate saving of at most 12.83\% and on average 10.24\%. BD-rate saving of 12.41\% and 14.24\%  for the chrominance component are obtained respectively. In the RA configuration, the YUV components obtain BD-rate saving on average of 3.57\% 5.38\% and 4.61\%, and the luminance component obtains at most 7.09\% BD-rate saving with sequence 13.

\begin{figure}[tbp]
\centering
\subfigure[HM Rec. 0.243bpp PSNR 37.90.]{
\begin{minipage}[t]{4.3cm}
\centering
\includegraphics[width=4.3cm]{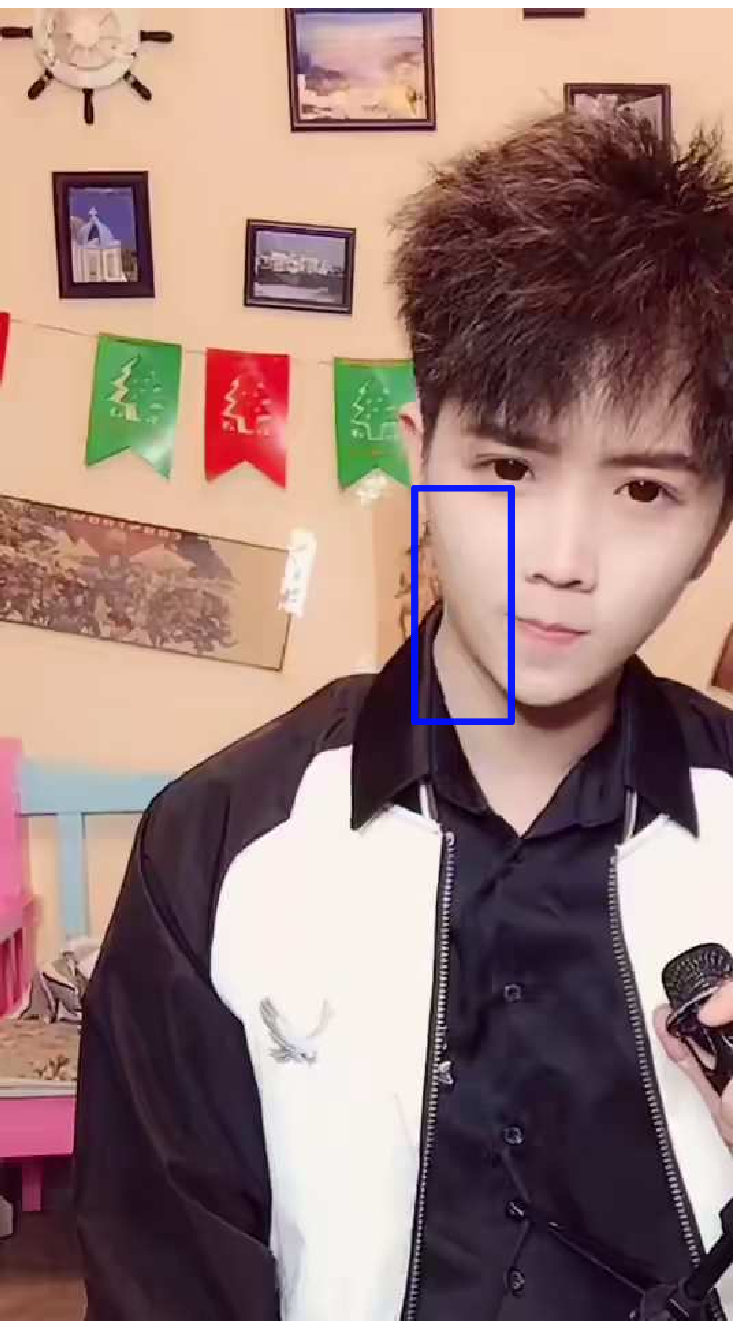}
\end{minipage}%
}%
\subfigure[Enhanced. 0.231bpp PSNR 38.31.]{
\begin{minipage}[t]{4.3cm}
\centering
\includegraphics[width=4.3cm]{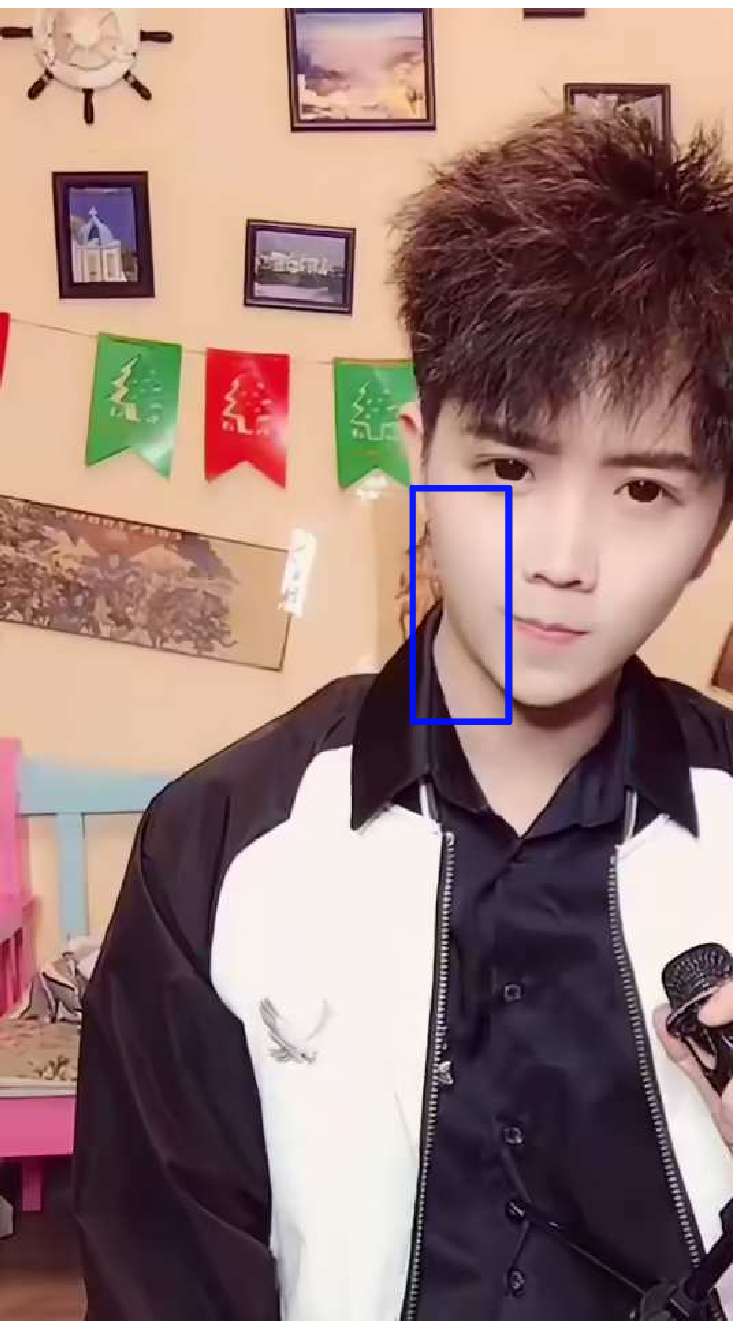}
\end{minipage}%
}%
\centering
\caption{Comparison between original reconstruction image and enhanced reconstruction image. This image comes from the 1-st frame in sequence 01 with QP is 32.}\label{sub}
\end{figure}

For the subjective image quality evaluation of the proposed method, Fig. \ref{sub} compares the HM reconstruction image (on the left) and the enhanced image by our networks (on the right). In the face area of the blue box, we can clearly see the contouring and blocky artifacts in Fig. \ref{sub}(a). On the other hand, in Fig. \ref{sub}(b), these artifacts are well eliminated and the face is smoother and plumper. Moreover, our model provides a higher compression ratio (0.231 bpp of our models to 0.243 bpp of HM).
\begin{table*}[!tbp]
\footnotesize
  \centering
  \caption{The comparison of several filtering neural network with HM baseline in All-Intra for PCS grand challenge short videos }\label{vrcnn}
  \begin{tabu}{c|r|r|r |r|r|r |r|r|r |r|r|r}
    \Xhline{1.0pt}
    \multirow{2}*{Sequence}  & \multicolumn{3}{c|}{ARCNN\cite{ARCNN}}& \multicolumn{3}{c|}{VRCNN\cite{VRCNN}}& \multicolumn{3}{c|}{Ours Filter.}& \multicolumn{3}{c}{Ours All.} \rule{0pt}{9pt}  \\[1pt]
    \cline{2-13}
    ~&\multicolumn{1}{c|}{Y}& \multicolumn{1}{c|}{U}&\multicolumn{1}{c|}{V}&\multicolumn{1}{c|}{Y}& \multicolumn{1}{c|}{U}&\multicolumn{1}{c|}{V}&\multicolumn{1}{c|}{Y}& \multicolumn{1}{c|}{U}&\multicolumn{1}{c|}{V}&\multicolumn{1}{c|}{Y}& \multicolumn{1}{c|}{U}&\multicolumn{1}{c}{V} \\
    \hline
    01 &-2.40\%&-4.24\%&-1.95\%&-3.65\%&-10.85\%&-6.51\%&-6.06\%&-12.51\%&-9.84\%&-8.63\%&-16.77\%&-13.57\%\\
    \hline
    02 &-2.43\%&-3.70\%&-2.99\%&-4.53\%&-6.69\%&-9.71\%&-8.49\%&-12.65\%&-14.58\%&-11.81\%&-19.18\%&-18.68\%\\
    \hline
    03 &-1.71\%&-1.93\%&-2.84\%&-3.07\%&-6.80\%&-4.18\%&-5.02\%&-8.56\%&-8.81\%&-7.19\%&-13.83\%&-15.96\%\\
    \hline
    06 &-3.04\%&-3.23\%&-2.77\%&-5.70\%&-6.50\%&-5.02\%&-9.67\%&-9.33\%&-6.14\%&-10.72\%&-12.49\%&-10.46\%\\
    \hline
    08 &-2.41\%&-3.29\%&-2.01\%&-4.26\%&-4.42\%&-3.49\%&-7.17\%&-6.76\%&-6.23\%&-9.31\%&-11.10\%&-9.94\%\\
    \hline
    09 &-1.13\%&-4.05\%&-2.47\%&-1.88\%&-10.55\%&-7.07\%&-3.19\%&-10.58\%&-9.65\%&-5.33\%&-14.84\%&-14.36\%\\
    \hline
    10 &-2.67\%&-0.50\%&-4.20\%&-4.95\%&-2.55\%&-7.42\%&-8.77\%&-8.73\%&-8.67\%&-11.05\%&-12.24\%&-14.16\%\\
    \hline
    13 &-2.58\%&-7.43\%&-1.79\%&-4.61\%&-15.63\%&-11.19\%&-8.81\%&-19.34\%&-16.01\%&-12.53\%&-23.42\%&-19.16\%\\
    \Xhline{0.8pt}
    Mean BD-rate &-2.30\%&-3.55\%&-2.63\%&-4.08\%&-8.00\%&-6.82\%&-7.15\%&-11.06\%&-9.99\%&\textbf{-9.57\%}&\textbf{-15.48\%}&\textbf{-14.54\%}\\
    \Xhline{1.0pt}
  \end{tabu}
\end{table*}


\subsection{Comparison with AR-CNN and VR-CNN}

We use the same dataset to train AR-CNN and VR-CNN and test them in the same situation. More specifically, in order not to bring in the influence of neural network-based intra prediction, the neural network prediction is turned off while testing the filtering networks. At the same time, we test them based on block-level filtering which may improve the reconstruction pixels so as to explore the potential of these filtering ways. All sequences are tested for the first frame under each QP, and the experimental results are shown in TABLE \ref{vrcnn}, we can find that.
\begin{itemize}
  \item Compared with the test results of our filter model ("Our Filter." column in the TABLE \ref{vrcnn}) used alone, using our joint model ("Our All." column in the TABLE \ref{vrcnn}) achieves better performance, the BD-rate of the YUV component is reduced by 2.42\%, 4.42\%, and 4.55\%. This means our combination of intra prediction and reconstruction filtering is successful.
  \item VR-CNN and our proposed method have better performance than AR-CNN because of the diversity of their convolution kernels in different subbranches. And they could get better performance in some sequences than others, such as 13. This may be related to the dataset used for training, because our dataset is a spliced sequence of static pictures (DIV2K), and those test sequences which have better performance are relatively static to some extent.
  \item Our filtering model has a stronger feature extraction capability than VR-CNN, and achieves up to 7.15\% and on average 9.67\% BD-rate saving. AR-CNN achieves 2.30\%, 3.55\% and 2.63\% and VR-CNN achieves 4.08\%, 8.00\% and 6.82\% respectively on the three components of YUV. Compared with VR-CNN and AR-CNN, our model has better performance in all sequences and components.
\end{itemize}

\subsection{Comparison with RHCNN}
In this subsection, we test the performance of our filtering model for the HEVC test sequences and compare it with RHCNN. It can be observed from the test results TABLE \ref{hmsd} that our filtering model also get results (9.70\%, 11.59\% and 13.35\% respectively on the three components of YUV) which close to the result of 8 provided sequences. And the dataset used for training doesn't overlap with the HEVC test sequences, thus further demonstrating the generalization ability of our model.
\begin{table}[tbp]
\small
  \centering
  \caption{BD-rate saving of proposed reconstruction filtering network than HM in All-Intra for HEVC test sequences}\label{hmsd}
  \begin{tabular}{l|l|r|r|r}
        \Xhline{1.0pt}
    \multirow{2}*{Class} & \multirow{2}*{Sequence} & \multicolumn{3}{c}{Our All.} \rule{0pt}{9pt}  \\
    \cline{3-5}
    ~&~&\multicolumn{1}{c|}{Y}&\multicolumn{1}{c|}{U}&\multicolumn{1}{c}{V}\\
    \hline
     \multirow{2}*{ClassA} & Traffic   & -11.68\%&-9.55\%&-11.51\%\\
    \cline{2-5}
    ~&PeopleOnStreet   & -11.45\%&-14.11\%&-14.00\%\\
    \hline
    \multirow{5}*{ClassB} &Kimono     & -5.97\%&-3.42\%&-3.06\% \\
    \cline{2-5}
    ~&ParkScene     & -8.65\%&-7.98\%&-7.43\% \\
    \cline{2-5}
    ~&Cactus    & -8.64\%&-10.33\%&-15.17\% \\
    \cline{2-5}
    ~&BasketballDrive     &-7.96\%&-11.56\%&-16.90\% \\
    \cline{2-5}
    ~&BQTerrace     &-6.61\%&-9.72\%&-10.92\% \\
    \hline
    \multirow{4}*{ClassC} &BasketballDrill      & -11.61\%&-16.27\%&-19.68\%\\
    \cline{2-5}
    ~  &BQMall      &-9.36\%&-11.77\%&-13.52\%\\
    \cline{2-5}
    ~  &PartyScene     & -6.08\%&-8.93\%&-9.97\% \\
    \cline{2-5}
    ~  &RaceHorses     & -6.91\%&-11.92\%&-17.54\% \\
    \hline
    \multirow{4}*{ClassD} &BasketballPass      & -9.05\%&-9.40\%&-12.48\%\\
    \cline{2-5}
    ~  &BQSquare     &-6.55\%&-5.33\%&-7.83\% \\
    \cline{2-5}
    ~  &BlowingBubbles     & -7.42\%&-11.24\%&-10.46\%\\
    \cline{2-5}
    ~  &RaceHorses  &-11.20\%&-15.11\%&-18.38\%\\
    \hline
    \multirow{6}*{ClassE} &Vidyo1 & -13.13\%&-13.85\%&-15.19\%\\
    \cline{2-5}
    ~  &Vidyo3  & -9.62\%&-7.07\%&-11.75\% \\
    \cline{2-5}
     ~  &Vidyo4   & -11.13\%&-14.72\%&-14.94\%\\
    \cline{2-5}
     ~  &FourPeople     & -14.90\%&-15.44\%&-16.76\% \\
    \cline{2-5}
     ~  &Johnny      & -12.95\%&-19.91\%&-16.10\%\\
    \cline{2-5}
     ~  &KristenAndSara    & -12.74\%&-15.73\%&-16.78\%\\
        \Xhline{0.8pt}

    \multicolumn{2}{c|}{Average} & -9.70\%&-11.59\%&-13.35\%\\
        \Xhline{1.0pt}
  \end{tabular}
\end{table}
\begin{table}[!tbp]
\small
  \centering
  \caption{Comparison of RHCNN and our reconstruction filtering network for All-Intra}\label{results}
  \begin{tabular}{l|c|c|c}
    \Xhline{1.0pt}
   Sequence & \multicolumn{1}{c|}{RHCNN\cite{RHCNN}} & \multicolumn{1}{c|}{Our Filter. }&
    \multicolumn{1}{c}{Our All.} \rule{0pt}{10pt}     \\[2pt]

    \hline
      Traffic  & -6.10\% & -7.72\% & \textbf{-11.40\%} \\
\hline
    PeopleOnStreet   &   -5.30\% & -7.80\% & \textbf{-11.75\%}\\
    \hline
    RaceHorses &   -5.60\%  & -8.37\% & \textbf{-12.04\%}\\
    \hline
    Vidyo1 &   -7.50\%& -8.59\%&\textbf{-13.27\%}\\
\hline
   Vidyo3 &  -6.40\% & -6.61\%&\textbf{-9.68\%} \\
\hline
  Vidyo4 &  -6.20\%  & -7.53\%&\textbf{-11.49\%}\\
    \Xhline{0.8pt}
    \multicolumn{1}{l|}{Mean BD-rate}     & -6.18\%  & -7.77\%& \textbf{-11.61\%}\\
    \Xhline{1.0pt}
  \end{tabular}
\end{table}

The trainable parameters number of our filter model is 475,233. At the same time, The RHCNN with 3,340,000 trainable parameters is used as a comparison. We compare the test results provided in their paper \cite{RHCNN} with our results. It can be seen that our filter models saving 7.77\% BD-rate on average and up to 11.61\% BD-rate saving is obtained by our joint model. Both results are better than RHCNN's performance while our models have fewer parameters.
\begin{table}[!tbp]
\small
  \centering
  \caption{Comparison of the number of model parameter}\label{para}
  \begin{tabular}{l|c|c|c|c}
    \Xhline{1.0pt}
   Model  & VR-CNN\cite{VRCNN}  &AR-CNN\cite{ARCNN}   & Our filter &RHCNN\cite{RHCNN}  \\
   \hline
   Number &54,512 &106,448& 475,233 &   3,340,000 \\
    \Xhline{1.0pt}
  \end{tabular}
\end{table}

\section{Conclusion}
In this paper, a  dual learning-based method in intra coding consists of intra prediction and reconstruction filtering was proposed for video coding.Inception dense blocks, which have strong feature extraction function is used for improving VR-CNN to make it preform better in reconstruction filtering. At the same time, we changed the frame-level filtering in previous works to block-level filtering, which can help reduce the extra bitstream caused by the error of the reference pixels. For PCS grand challenge short videos test sequences, experimental results show that our networks achieve on average 10.24\% BD-rate saving for all-Intra and 3.57\% BD-rate saving for random -access.
Our model also obtains a 9.70\% BD-rate saving compared to HM-16.20 baseline for HEVC test sequences.
Finally, we compared our reconstruction filtering model with many other models, and our models achieved the best performance.  
In future work, we will continue to explore the more efficient and light models to achieve better coding performance.

\section*{Acknowledgment}
This work was supported in part by the National Natural Science Foundation of China under Grant 61674041, in part by Alibaba Group through Alibaba Innovative Research (AIR) Program, in part by the STCSM under Grant 16XD1400300, in part by the pioneering project of academy for engineering and technology and Fudan-CIOMP joint fund.

This work was supported by the National Natural Science Foundation of China under Grant 61525401，the Program of Shanghai Academic/Technology Research Leader under Grant 16XD1400300, the Innovation Program of Shanghai Municipal Education Commission.

\bibliography{references}{}
\bibliographystyle{IEEEtran}
\end{document}